# Thermal expansion of α-boron and some boron-rich pnictides

Kirill A. Cherednichenko and Vladimir L. Solozhenko [*]

*LSPM-CNRS, Université Paris Nord, 93430 Villetaneuse, France*

**Abstract**

*Thermal expansion of α-rhombohedral boron (α-$B_{12}$) and two isostructural boron-rich pnictides ($B_{12}P_2$ and $B_{12}As_2$) has been studied between 298 and 1280 K by high-temperature synchrotron X-ray diffraction. For all studied phases no temperature-induced phase transitions have been observed. The observed temperature dependencies of the lattice parameters and unit cell volumes were found to be quasi-linear. Variation of the thermal expansion coefficients in the group of boron-rich pnictides ($B_{13}N_2$ – $B_{12}P_2$ – $B_{12}As_2$) was analyzed.*

## I. Introduction

During last few decades elemental boron and boron-rich compounds have been attracting a considerable attention due to their unusual crystal structure and properties. The common feature of these compounds is $B_{12}$ icosahedral *closo*-clusters with polycentric metal-like bonding system [1]. At the same time, boron atoms within icosahedra form strong covalent bonds with neighbouring atoms of the adjacent icosahedra and various interstitial atoms (e.g. C, N, O, P, etc.). A combination of the metal-like intra-icosahedral bonds and strong covalent inter-icosahedral bonding makes boron-rich compounds extremely stable, eventually leading to high melting temperatures, chemical inertness, outstanding mechanical properties and interesting electronic properties (e.g. self-healing resistance to radiation damage, etc.) [1-7]. It should be noted that properties of boron-rich solids are significantly influenced by the interstitial atoms [3,7]. Thus, study of the impact of interstitial atoms on various properties (hardness, compressibility, thermal expansion, band gap, etc.) is of great interest from both fundamental and application points of view.

Very recently thermal expansion of two boron-rich nitrides synthesized under extreme conditions has been studied by synchrotron X-ray diffraction [8]. In the present paper we report the results on thermal expansion of α-rhombohedral boron (α-$B_{12}$) and two boron-rich pnictides, $B_{12}P_2$ and $B_{12}As_2$. The retrieved thermal expansion coefficients compared with those reported earlier [8,9] allowed us to analyze the thermal expansion variation in the family of boron-rich pnictides.

[*] Corresponding author: vladimir.solozhenko@univ-paris13.fr

## II. Experimental

Highly crystalline α-B$_{12}$ (99.98%) was provided by Dr. Igor N. Goncharenko, Laboratoire Léon Brillouin (CEA-Saclay, France). Boron subphosphide B$_{12}$P$_2$ was produced by self-propagating high-temperature reaction in the BPO$_4$–MgB2–Mg system according to the method described elsewhere [10]. Boron subarsenide B$_{12}$As$_2$ was synthesized by reaction of amorphous boron with arsenic melt at 5 GPa and 2100 K in a toroid-type apparatus (for details, see [11]). According to X-ray diffraction study (TEXT 3000 Inel diffractometer, CuKα1 radiation) all samples are single-phase, and their lattice parameters determined using Maud [12] and PowderCell [13] software are in good agreement with literature data [1,3].

Thermal expansion of boron-rich pnictides and α-B$_{12}$ has been studied at MCX beamline, Elettra Synchrotron (Trieste, Italy) and B2 beamline, DORIS III storage ring, HASYLAB-DESY (Hamburg, Germany). The powdered samples were loaded into quartz-glass capillaries under Ar atmosphere. Debye-Scherrer geometry with rotating capillary was used at both beamlines. The wavelength (1.0352 Å at MCX beamline and 0.6841 Å at B2 beamline) was calibrated using Si and LaB$_6$ as external standards; temperature calibration has been done by measuring thermal expansion of platinum [14] under the same experimental conditions. X-ray diffraction patterns were collected in the 2-35 θ-range using image plate detectors upon stepwise heating with 100-K steps from room temperature up to 1280 K. The ramp time was 10 min with 5 min of temperature stabilization; the acquisition time was 60-120 s. FIT2D software [15] was used to integrate the collected X-ray diffraction data into one-dimensional patterns. The lattice parameters of the studied phases at different temperatures were determined by Le Bail refinement procedure using PowderCell software [13].

## III. Results and Discussion

α-rhombohedral boron is the low-temperature allotrope, and its thermodynamic stability domain in the equilibrium phase diagram of boron [16] is still not precisely defined. It has rhombohedral crystal structure (*R-3m*) containing B$_{12}$ icosahedra located at the vertices of the rhombohedral cell. Due to the lack of the valence electrons necessary for the formation of inter-icosahedral covalent bonds, the equatorial boron atoms of the neighboring icosahedra lying in one layer forming relatively weak "two electrons-three centers" bonds, *2e3c* [1,2] (Fig. 1a). The boron-rich pnictides, B$_{12}$X$_2$Y (where X = N, P, As; Y = B), have structures related to α-rhombohedral boron with two or three interstitial atoms in 1*b* and 2*c* Wyckoff positions (in hexagonal setting, 3*b* and 6*c*) placed along the body diagonal of the rhombohedral unit cell (Fig. 1b). The interstitial atoms in 2*c* positions form the "two electrons-two centers" covalent bonds, *2e2c*, with equatorial boron atoms of the neighboring icosahedra.

The unit cell parameters of α-B$_{12}$, B$_{13}$N$_2$ [8], B$_{12}$P$_2$ and B$_{12}$As$_2$ in hexagonal ($a^{hex}$, $c^{hex}$) and rhombohedral ($a^{rh}$, $α^{rh}$) settings are presented in Table I. Further we will consider the thermal expansion in hexagonal unit cell only, so the "*hex*" indexes of lattice parameters will not be used. As one can see, *a* parameter increases in the α-B$_{12}$ – B$_{13}$N$_2$ – B$_{12}$P$_2$ – B$_{12}$As$_2$ row, which happens due to insertion of the interstitial atoms in α-boron unit cell and subsequent "squeezing" of the B$_{12}$ units outside [3]. On the other hand, X-X bonds (X = N, P, As) tend to pull closer the adjacent planes of icosahedra resulting in decrease of $c$ parameter [17]. Different "compression" of boron-rich pnictides along $c$-axis might be explained by different nature of X-X bonds i.e. P–P and As–As are *2e2c* covalent bonds (however, they are considerably longer than the corresponding diatomic equilibrium bonds [18], see Table II), whereas B–N–B chain in the $B_{13}N_2$ structure can be considered as *3e3c* bond (for details, see [19]), and is significantly longer. Nevertheless, as it follows from the formula for the hexagonal unit cell volume:

$$V_0 = \frac{\sqrt{3}}{2} a^2 c \qquad (1)$$

the enlargement of hexagonal unit cell in the $ab$-plane has significantly greater impact on increase of the unit cell volume than the "compression" along $c$-axis.

The results of high-temperature X-ray diffraction study of α-$B_{12}$, $B_{12}P_2$ and $B_{12}As_2$ (Fig. 2) revealed quasi-linear temperature dependency of the lattice parameters, hence, the linear thermal expansion coefficients (TEC) do not change over the whole temperature range under study. The linear TECs ($α_l$) can be estimated by Eq. 2:

$$\alpha_l = \frac{l - l_{298\,\text{K}}}{l_{298\,\text{K}} \cdot (T - 298\,\text{K})} \qquad (2)$$

where $l$ is a unit cell parameter and $l_{298\,K}$ is a unit cell parameter at room temperature. The $α_l$ values are presented in Table I. The thermal expansion of different interatomic distances was not analyzed due to insufficient quality of powder X-ray diffraction data (for that the single-crystal diffraction data are highly required); here and further the thermal expansion of the inter- and intra-icosahedral bonds is assumed to be the same for all studied solids.

As it has been expected, thermal expansion of α-rhombohedral boron and isostructural boron-rich pnictides in different crystallographic directions was found to be anisotropic. The anisotropy was estimated by $α_c/α_a$ ratio (see Table I): in contrast to α-$B_{12}$, thermal expansion of all boron-rich pnictides along $c$-axis ($α_c$) is larger than that in the $ab$-plane ($α_a$) (see Fig. 3$a$). Moreover, the liner TEC values of α-boron and boron-rich pnictides vary considerably.

Analysis of the $α_c$ variation of boron-rich pnictides requires special consideration of B–X and X–X bonds. According to Fig. 1, the interstitial atoms (X = N, P, As) in $3c$ Wyckoff positions form the bonds with four neighboring atoms and, hence, are tetrahedrally coordinated as well as X atoms in cubic boron pnictides BN [20], BP [21] and BAs [22]. All four B–X bonds in the cubic BX phase have equal lengths, thus, forming an ideal tetrahedron. Taking into account thermodynamic stability of all cubic boron pnictides, the lengths of the B–X bonds in these compounds were assumed to be the optimal for the tetrahedral geometry. Unlike cubic boron pnictides, the tetrahedra (with central X atom) in $B_{13}N_2$, $B_{12}P_2$ and $B_{12}As_2$ are significantly distorted (see Table II). Employing the explanation that has been already used in the case of $B_{50}N_2$ thermal expansion [8] we assume that thermal vibrations lead to the reduction of the distorted tetrahedra towards the ideal ones i.e. at high temperatures the X–X bonds in $B_{12}P_2$ and $B_{12}As_2$ tend to shrink (see Table II for corresponding diatomic equilibrium bonds), whereas N–B–N chains in $B_{13}N_2$ tend to expand. In the case of boron subnitride, nitrogen atom is in the center of tetrahedron formed by boron atoms only, that is why the B–N distances in $B_{13}N_2$ are comparable with the corresponding bond lengths in cubic BN, but not with diatomic equilibrium B–N bond. Such a different thermal expansion of boron-rich pnictides along $c$-axis results in $α_c$ decrease in the row: $B_{13}N_2$ – $B_{12}P_2$ – $B_{12}As_2$ (see Fig. 3$a$). Despite one could expect the maximal thermal expansion in *c*-direction for α-$B_{12}$, its $α_c$ value was found to be close to that of $B_{12}As_2$.

Following the same logic, one should expect that at high temperatures B–P and B–As bonds in the *ab*-plane will tend to expand (see the difference between B–X bond lengths in $B_{12}X_2$ and cubic BX, Table II), while B–N bonds will tend to shrink, which should result in the higher $α_a$ values for $B_{12}P_2$ and $B_{12}As_2$. Nevertheless, the hexagonal unit cell initially enlarged in the *ab*-plane leads to the smaller increase of the inter-icosahedral distances and B–X bonds at high temperatures, and, thus, to lower $α_a$ values; $α_a$ decrease in the row $B_{13}N_2$ – $B_{12}P_2$ – $B_{12}As_2$. The largest thermal expansion of α-$B_{12}$ in the *ab*-plane might be explained by the presence of weak *2e3c* inter-icosahedral bonds instead of *2e2c* covalent bonds in boron-rich pnictides i.e. stronger *2e2c* bonds prevent significant thermal expansion in this crystallographic direction.

Fig. 2 presents the variations of the normalized unit cell volumes $V(T)/V_0$ ($V_0$ is the unit cell volume at 298 K) of α-$B_{12}$, $B_{12}P_2$ and $B_{12}As_2$ *versus* temperature. The observed temperature dependencies of unit cell volume are quasi-linear and, thus, can be approximated by Eq. 3:

$$V(T) = V_0\big(1 + α_v(T - 298 \text{ K})\big), \tag{3}$$

where $α_v$ is volume TEC.

The retrieved volume TECs of α-$B_{12}$, $B_{12}P_2$ and $B_{12}As_2$ (see Table I) were found to be of the same order of magnitude as those of other boron-rich solids [8,9,23-28]. The $α_v$ value of $B_{12}As_2$ (15.3(1)×$10^{-6}$ $K^{-1}$) is in excellent agreement with the literature data (15(2)×$10^{-6}$ $K^{-1}$) [9]. According to Table I, boron subnitride has the highest volume TEC among all studied phases. The main reasons for that are: not too much expanded unit cell at room temperature (thus, allowing the further increase of the interatomic distances) and the presence of *3e3c* N–B–N chains with its tendency to expand at high temperatures. α-$B_{12}$ has the second largest volume TEC owing to strong impact of the rather weak *2e3c* bonds resulting in significant expansion in the *ab*-plane. The volume thermal expansion of boron subphosphide and subarsenide is considerably influenced by the "negative" tendency for the P–P and As–As bonds at high temperatures and the unit cells initially enlarged in *a* and *b* directions.

Interestingly, the volume TECs and bulk moduli ($B_0$) of $B_{13}N_2$, $B_{12}P_2$ and $B_{12}As_2$ [29-31] vary similar (see Table I and Fig. 3*b*). As one can see, the $α_v$ *vs* $B_0$ dependence for boron-rich pnictides can be approximated by the following exponential function:

$$α_v = 14.94 + 1.37 \cdot 10^{-4} \cdot e^{0.05 \cdot B_0}, \tag{4}$$

Data for α-rhombohedral boron is also presented in Fig. 3*b* (we used $B_0$ value of α-$B_{12}$ reported in [32]), however it does not follow the exponential $α_v(B_0)$ dependency. It should be noted that for all studied phases we did not observe temperature-induced phase transitions and/or decomposition up to the highest experimental temperatures.

IV. Conclusions

Thermal expansion of α-rhombohedral boron and isostructural boron-rich pnictides ($B_{12}P_2$, $B_{12}As_2$) was studied *in situ* by synchrotron X-ray diffraction up to 1280 K. The precise measurements of lattice parameters at different temperatures allowed us to retrieve the corresponding linear ($α_l$) and volume ($α_v$) thermal expansion coefficients of α-$B_{12}$, $B_{12}P_2$ and $B_{12}As_2$. The obtained values were compared with the literature data for boron subnitride $B_{13}N_2$, and variation of $α_l$ and $α_v$ of three boron-rich pnictides was analyzed in terms of their crystal structure.

## Acknowledgements

The authors thank Dr. Vladimir Mukhanov for the samples synthesis and Drs. Jasper Rikkert Plaisier, Lara Gigli and Giulio Zerauschek for assistance in measurements at Elettra. Synchrotron X-ray diffraction experiments were carried out during beam time allocated for Proposal 20160086 at Elettra Sincrotrone Trieste and Proposal I-20090172 EC at HASYLAB-DESY. This work was financially supported by the European Union's Horizon 2020 Research and Innovation Programme under Flintstone2020 project (grant agreement No 689279).

## ORCID IDs

Vladimir L. Solozhenko  https://orcid.org/0000-0002-0881-9761

**Table I.** Lattice parameters, unit cell volumes, mean linear thermal expansion coefficients ($\alpha_l$), linear thermal expansion anisotropy parameters ($\alpha_a/\alpha_c$), volume thermal expansion coefficients ($\alpha_v$) and bulk moduli ($B_0$) of α-$B_{12}$, $B_{13}N_2$, $B_{12}P_2$ and $B_{12}As_2$.

|  | α-$B_{12}$ | $B_{13}N_2$ [8] | $B_{12}P_2$ | $B_{12}As_2$ |
|---|---|---|---|---|
| $a,b^{hex}$ Å | 4.9161(3) | 5.4537(3) | 5.9894(3) | 6.1353(2) |
| $c^{hex}$, Å | 12.5752(5) | 12.2537(7) | 11.8594(8) | 11.8940(7) |
| $V^{hex}$, Å$^3$ | 263.20(4) | 315.62(5) | 368.43(6) | 387.72(5) |
| $a^{rh}$, Å | 5.0623(3) | 5.1573(3) | 5.2521(1) | 5.3166(2) |
| $\alpha^{rh}$, ° | 58.09 | 63.84 | 69.53 | 70.48 |
| $V^{rh}$, Å$^3$ | 87.73(1) | 105.21(2) | 122.81(2) | 129.24(2) |
| $\alpha_a \times 10^6$, K$^{-1}$ | 6.4(3) | 6.1(1) | 5.7(1) | 5.0(1) |
| $\alpha_c \times 10^6$, K$^{-1}$ | 5.4(2) | 8.9(1) | 6.5(1) | 5.3(1) |
| $\alpha_c/\alpha_a$ | 0.84 | 1.46 | 1.14 | 1.06 |
| $\alpha_v \times 10^6$, K$^{-1}$ | 18.3(6) | 21.3(2) | 17.9(3) | 15.3(1) |
| $B_0$, GPa | 224(7) [32] | 205(2) [31] | 192(11) [30] | 150(4) [29] |

**Table II.** Bond lengths in boron-rich pnictides and cubic boron pnictides compared with diatomic equilibrium B–B and B–X (X = N, P, As) distances.

|  | BN [20] | BP [21] | BAs [22] | α-$B_{12}$ | $B_{13}N_2$ [8] | $B_{12}P_2$ | $B_{12}As_2$ |
|---|---|---|---|---|---|---|---|
| B–X, Å | 1.566 | 1.967 | 2.069 | – | 1.6302 | 1.9074 | 1.9914 |
| X–X, Å | – | – | – | – | 1.5390* | 2.2428 | 2.3833 |
| Diatomic equilibrium bonds [18], Å | | | | B–B | B–N | P–P | As–As |
| Diatomic equilibrium bonds [18], Å | | | | 1.590 | 1.281 | 1.893 | 2.103 |

\* X–X distance in $B_{13}N_2$ is given for the N–B bond of the N–B–N chain.

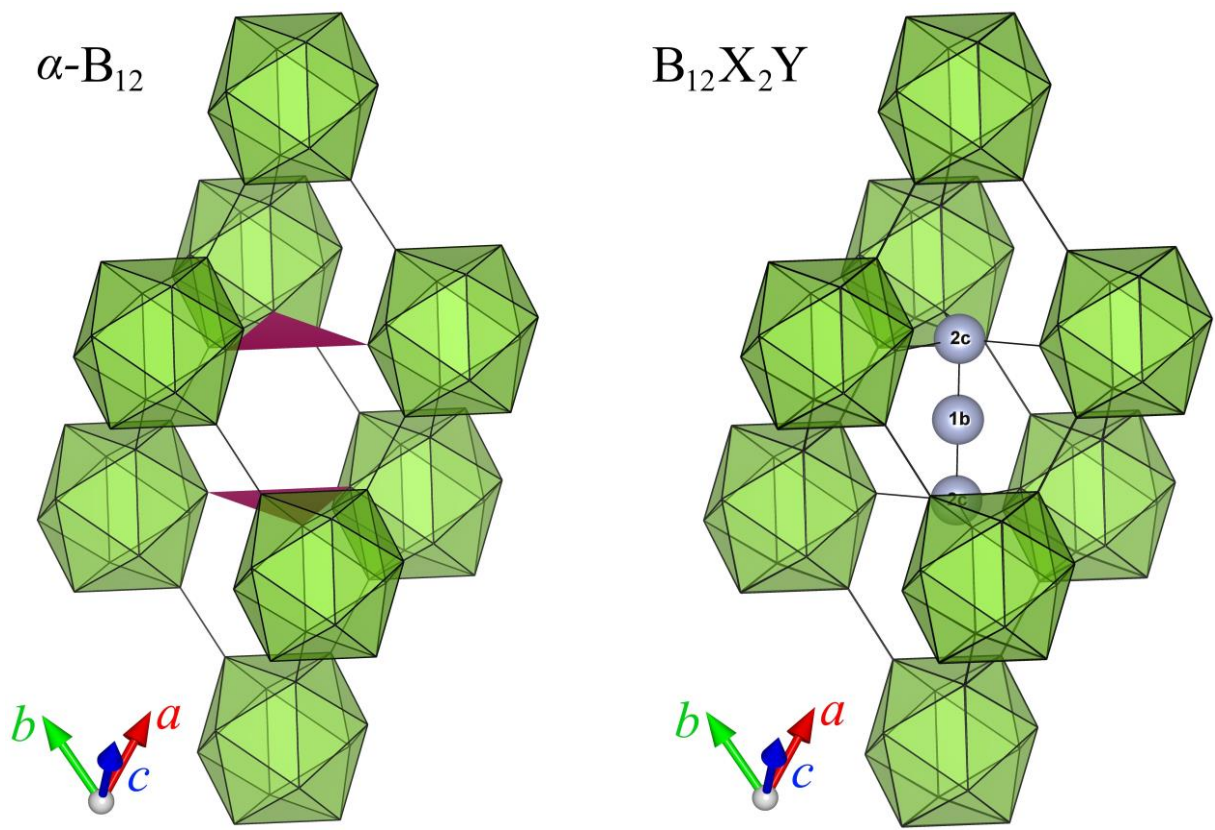

Fig. 1. Crystal structures of α-rhombohedral boron (*left*) and α-B$_{12}$-related boron-rich pnictides, B$_{12}$X$_2$Y, (*right*) in rhombohedral setting. The B$_{12}$ clusters are presented by green icosahedra, the *2e3c* bonds in α-B$_{12}$ are shown by rose triangles, the interstitial X (X = N, P, As) and Y atoms (Y = B) in *2c* and *1b* Wyckoff positions, respectively, are shown by grey balls.

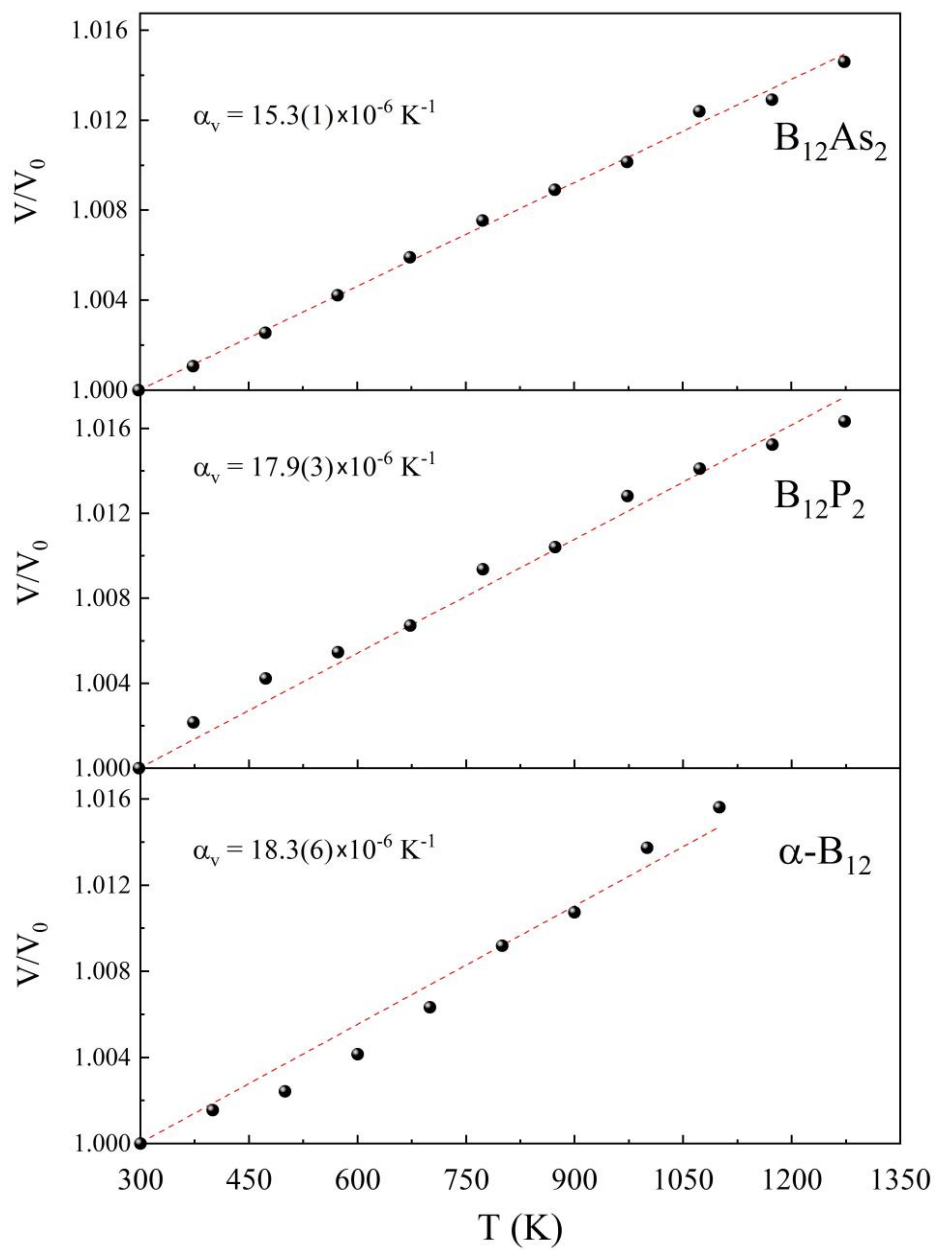

Fig. 2. Variation of the normalized unit cell volumes of α-$B_{12}$, $B_{12}P_2$ and $B_{12}As_2$ as function of temperature. The dashed lines represent the linear fits to the experimental data. The corresponding values of volume thermal expansion coefficients are indicated.

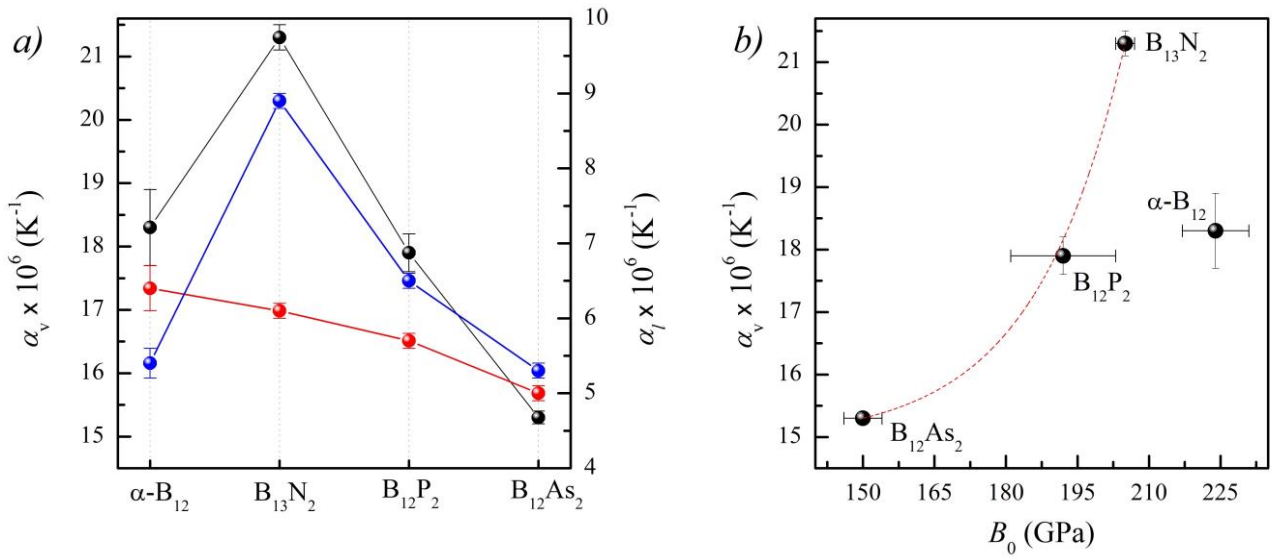

Fig. 3. ***a***) Volume ($α_v$, black) and linear ($α_a$, red, and $α_c$, blue) thermal expansion coefficients of α-$B_{12}$, $B_{13}N_2$ [8], $B_{12}P_2$ and $B_{12}As_2$; ***b***) Volume ($α_v$) thermal expansion coefficients of α-$B_{12}$, $B_{13}N_2$, $B_{12}P_2$, $B_{12}As_2$ *versus* bulk modulus ($B_0$) [29-32]; the dashed line represents exponential fit.